\documentclass[12pt]{article} 
\usepackage{cyr}
\usepackage{epsf}
\usepackage{pazha}
\tightenlines

\def\*{$^{*}$}
\def\Á{$^{\mbox{\small a}}$}
\def\Â{$^{\mbox{\small b}}$}
\def\×{$^{\mbox{\small c}}$}
\def\Ç{$^{\mbox{\small d}}$}
\def\Ä{$^{\mbox{\small e}}$}

\sloppypar

\begin{document}
\baselineskip 21pt

\title{\bf RELATIVE CONTRIBUTION OF THE HYDROGEN 2S TWO-PHOTON DECAY AND
  LYMAN-$\alpha$ ESCAPE CHANNELS DURING THE EPOCH OF COSMOLOGICAL RECOMBINATION }

\author{\bf \hspace{-1.3cm}\copyright\, 2017 \ \ 
J. A. Rubi\~no-Martin\affilmark{1,2*}, R. A. Sunyaev\affilmark{3,4}}

\affil{
{\it Instituto de Astrofisica de Canarias, La Laguna, Tenerife, Spain}$^1$\\
{\it Departamento de Astrofisica, Universidad de La Laguna,  La Laguna, Tenerife, Spain}$^2$\\
{\it Max-Planck-Institut fuer Astrophysik, Garching, Germany}$^3$\\
{\it Space Research Institute (IKI), Russian Academy of Sciences,
  Moscow, Russia}$^4$ }

\vspace{2mm}
\received{\today}

\sloppypar 
\vspace{2mm}
\noindent

We discuss the evolution of the ratio in number
of recombinations due to 2s two photon escape and due to the escape of
Lyman-$\alpha$ photons from the resonance during the epoch of
cosmological recombination, within the width of the last scattering surface and near its boundaries.
We discuss how this ratio evolves in time, and how it defines the profile of the 
Lyman-$\alpha$ line in the spectrum of CMB. One of the key reasons for
explaining its time dependence is the strong overpopulation of
the 2p level relative to the 2s level at redshifts $z \la 750$. 


\noindent
{\bf PACS codes:\/} ?????

\vfill
\noindent\rule{8cm}{1pt}\\
{$^*$ E-mail $<$jalberto@iac.es$>$}

\clearpage

\section*{INTRODUCTION}
\noindent

During the last two decades, detailed observations of the
angular fluctuations of the cosmic microwave background (CMB) have
provided a very detailed description of the global properties of our
Universe (Hinshaw et al., 2013; Planck Collaboration XVI 2014; Planck
Collaboration XIII 2016). 
One of the key ingredients which are required to extract those
cosmological predictions from the CMB power spectra is an accurate description of
the background ionisation history in the Universe, and in particular,
of the process of cosmological recombination which takes the Universe from a state of
fully ionised hydrogen and helium at early times, to an almost
neutral medium ($x_{\rm e} \sim 2 \times 10^{-4}$) during the dark ages. 
This cosmological recombination occurring at redshift $z \sim
1100$ (Zeldovich, Kurt \& Sunyaev 1968; Peebles 1968; Seager, Sasselov
\& Scott  2000) also produces small spectral distortions of the CMB blackbody spectrum arising as a
consequence of the non-equilibrium conditions involved in that
process (Zeldovich, Kurt \& Syunyaev 1968; Dubrovich 1975; Sunyaev \&
Chluba 2009). This cosmological recombination spectrum has been described in detail in
the literature (Rubi\~no-Martin et al., 2006; Chluba \& Sunyaev 2006a; Chluba et al., 2010; Ali-Haimoud
2013; Chluba \& Ali-Haimoud 2016). 

Since the seminal papers of the late 1960s on the hydrogen
recombination problem (Zeldovich, Kurt \& Sunyaev 1968; Peebles 1968),
it is well-known that the 2s-1s two-photon decay channel plays a
fundamental role in controlling the dynamics and the timescales of the
process. 
The slow escape of photons from the Lyman-$\alpha$ resonance
strongly suppresses the effective rate of recombinations via the the
2p-1s channel, making the 2s-1s two-photon decay rate ($A_{2s,1s}
\approx 8.22$\,s$^{-1}$) the dominant route for recombination.  
In total, about 57\,\% of all hydrogen atoms in the Universe became
neutral through the 2s−1s channel, while the remaining 43\,\% used the
2p-1s route (e.g., Chluba \& Sunyaev 2006a; Wong et al. 2006).

As a historical note, the escape rate for cosmological Lyman-$\alpha$
line in a cosmological environment was derived by Zeldovich, Kurt \&
Sunyaev (1968, hereafter ZKS68) for cosmological recombination, and by 
Varshalovich and Syunyaev (1968) for the period after cosmological
reionization. 
The ZKS68 result was formally similar to the result of Sobolev (1947, 1960) for the expanding envelopes of stars (which is a
coordinate dependent problem with no dependence on time but with a
boundary conditions). It is a pity, but authors (and the referees)
of those papers mentioned above did not point out the analogy between the
problem in the cosmological context and the classical result of
V.V. Sobolev, and did not try to proof the
correspondence of the results of two different problems: time
dependent (cosmological) and steady state (expanding envelopes).
This analogy was later pointed out by e.g. Rybicki \& Dell'antonio (1993). 

Here we revisit the problem of escape of photons during recombination, and we explicitly present the relative
importance of both channels (2s-1s and 2p-1s) during the width of the
last scattering surface, but also at later times. Our main result, shown in
Figure~\ref{fig:frates}, is that the 2s-1s channel is only effectively
dominant during the redshift interval $680 \la z \la 1400$. At lower
redshifts, during the dark ages, the Lyman-$\alpha$ channel becomes
more important again. 

As we show below, the explanation for this behaviour is the following. 
At redshifts higher than $z\sim 1200$, the degree of ionisation was
very high, effectively decreasing the optical depth in the center of the
Lyman-$\alpha$ line. As a result, the role of Lyman-$\alpha$ escape
was more important than the 2s two-photon decay. 
At redshifts smaller than $z\sim 1100$, the degree of ionisation was
rapidly decreasing. Simultaneously, the temperature of
the radiation and the amount of photons able to ionize 2p and 2s
levels were also decreasing. These two facts, together with the dynamics of the
recombination process (the cascade of electrons from high levels),
produce a deviation of the ratio of the populations of 2s and 2p from statistical
equilibrium. In particular, at late times we find an over-population of the 2p over
the 2s level, and therefore the Lyman-$\alpha$ channel becomes
dominant again.

\section*{RESULTS AND DISCUSSION}
\noindent

As described in ZKS68 and Peebles (1968), the dynamics of the cosmological
hydrogen recombination is dictated by the  2s-1s two-photon decay channel.
Indeed, this channel defines the full shape (and in particular the
width and low redshift wings) of the Thomson visibility function
(Sunyaev \& Zeldovich 1970), computed as $\mathcal{V}_T(\eta)  =
(d\tau_T/d\eta) \exp(-\tau_T)$, and where $\tau_T$ is the Thomson
optical depth and $\eta$ is the conformal time.  This function has a
peak around $z\sim 1100$.  
In particular, according to the recent measurements of the cosmological parameters by the {\it PLANCK} satellite,
the redshift of last scattering (defined so that the optical depth to
Thomson scattering equals unity) is $z_\star=1089$ (e.g. Planck
Collaboration XIII 2016). 

Indirectly, the visibility function also defines the shape and the
frequency dependence of the fluctuations imprinted in the angular
power spectrum of the CMB, both in intensity (Rubi\~no-Martin,
Hernandez-Monteagudo \& Sunyaev 2005) and polarization
(Hernandez-Monteagudo, Rubi\~no-Martin \& Sunyaev 2007). 

Moreover, one can also use the high sensitivity of the recombination
process to the 2s-1s two-photon transition rate as a way of directly
measuring the corresponding Einstein-A coefficient from CMB data
(see e.g. Chluba \& Sunyaev 2006b, and Mukhanov et al., 2012). 
This has been done in Planck Collaboration XIII (2016), which provides a
determination of $A_{2s,1s}$ with an 8 per cent error ($7.75 \pm
0.61$\,s$^{-1}$), and a value which is fully consistent with the theoretical computation of 8.2206\,s$^{-1}$ from
Labzowsky et al. (2005).

\subsection*{Redshift of formation of the recombination lines}
\noindent
As discussed in Rubino-Martin et al., (2006, hereafter RCS06), the
redshift of formation of the hydrogen recombination lines is higher
than the peak of the visibility function, so in principle future
observations of these lines should allow us to explore earlier epochs of
the Universe than those explored by CMB anisotropies, where the optical depth due to Thomson scattering is
very high.  

The bulk of the Lyman-$\alpha$ emission originates at redshifts of $z
\approx 1400$, while the higher level transitions (e.g. H-$\alpha$)
peak at slightly lower redshifts ($z \approx 1300$). 
It is interesting to point out that the profile of these recombination lines is affected by the  2s-1s
two-photon decay rate. Figure~\ref{fig:zformation} shows the (normalised) intensity of
the Lyman-$\alpha$ and H-$\alpha$ lines as a function of redshift. The
figure is adapted from Fig.~9 in RCS06.

We note that for the H-$\alpha$ line, both recombinational channels (2s decay and Lyman-$\alpha$ escape)
contribute to its brightness, while for the the Lyman-$\alpha$ line,
only one single mechanism operates (Lyman-$\alpha$ escape).  
Indeed, the Lyman-$\alpha$ low frequency wing is suppressed due to the presence of the 2s decay. 
This contributes to the shift of the peak of the Lyman-$\alpha$ line towards higher
frequencies. 
The effect is also clearly seen when comparing the effective rates for both
transitions, as shown below.

\subsection*{Relative contribution of the 2s-1s and the 2p-1s channels}
\noindent
We now quantify  the relative weight of the 2s two-photon decay 
channel when compared with the standard Lyman-$\alpha$ channel, 
not only during the peak of the visibility function (around $z\sim
1100$), but also before and after this epoch. 
Again, all the results included in this subsection follow from the computations
carried out with the code presented in RCS06.  

The two main quantities to be compared in this paper are the effective
radiative rates for the 2s two-photon decay and Lyman-$\alpha$
channels, namely $\Delta R(2s,1s)$ and $\Delta R(2p,1s)$ respectively.
Following RCS06, the effective radiative rate for any given
transition from and upper level $i$ to a lower level $j$ is given
by\footnote{See RCS06 for a description of the assumptions 
  involved in this expresion, and the details on the definition of the 
  Sobolev escape probability. }
\begin{equation}
\label{eq:DRij}
\Delta R_{i,j}=p_{i,j}\,\frac{A_{i,j}\, N_i\,e^{h\nu_{i,j}/k
    T_\gamma}}{e^{h\nu_{i,j}/k T_\gamma}-1}
\left[1-\frac{g_i}{g_j}\,\frac{N_j}{N_i}\,e^{-h\nu_{i,j}/k T_\gamma}\right],
\end{equation}
where $T_\gamma$ is the temperature of the ambient photon field,
$\nu_{i,j}$ is the transition frequency, $A_{i,j}$ is the corresponding
Einstein $A$-coefficient, $p_{i,j}$ is the Sobolev escape
probability, and $N_i$ ($N_j$) and $g_i$ ($g_j$)
are the population and statistical weight of the upper (lower) level, respectively. In our
particular case, $j$ corresponds to the 1s level, and $i$
can be either 2s or 2p.

We note here that the approach of ZKS68, which was based on the radiative
transfer equation around the Lyman-$\alpha$ line, is mathematically
equivalent to the previous detailed-balance equation~\ref{eq:DRij} using the Sobolev escape
probability in the limit of very high optical depth $\tau_S$, and assuming
that we have complete redistribution of the emitted photons, and equal
absorption and emission profiles. Here, we have
\begin{equation}
\tau_S = \frac{3 A_{2p,1s} \lambda_\alpha^3}{8\pi H(z)} [N(1s) - N(2p)/3].
\end{equation}
The approach of ZKS68 was based on the equation of radiative
transfer. Using the method of characteristics for solving their partial
differential equation (see appendix in ZKS68), the solution gives the
rate of photon flow along the characteristics due to the Hubble
expansion of the Universe. Recently, the same approach was used by Ali-Haimoud
et al., (2010), who in addition confirmed that the result of this
simple approach gives a result which is exactly the same as the one
from the Sobolev approximation. We also note that the ZKS68 solution is valid for any time
dependence of the Hubble constant parameter $H(z)$. A detailed
  discussion on the validity of the Sobolev approximation for the
  cosmological recombination process can be found in Chluba \& Sunyaev (2009a,b). 

Figure~\ref{fig:rates} shows the redshift dependence of the net radiative
rates for both channels. It is adapted from Fig.~10 in RCS06, by
extending the redshift range down to $z=600$, and representing 
the vertical axis in logarithmic coordinates. 
As already discussed elsewhere, the 2s-1s rate dominates during the epoch of
recombination, while the escape through the Lyman-$\alpha$ dominates
at earlier times (above $z\sim 1400)$. 
However, in the low redshift tail of the visibility function, we can
see that at redshifts below $z \sim 680$, the  Lyman-$\alpha$ escape
route starts to dominate again. This issue was not discussed in
RCS06, so we present the discussion for the first time here. 

\begin{figure}[h]
\epsfxsize=19cm
\hspace{-2cm}\epsffile{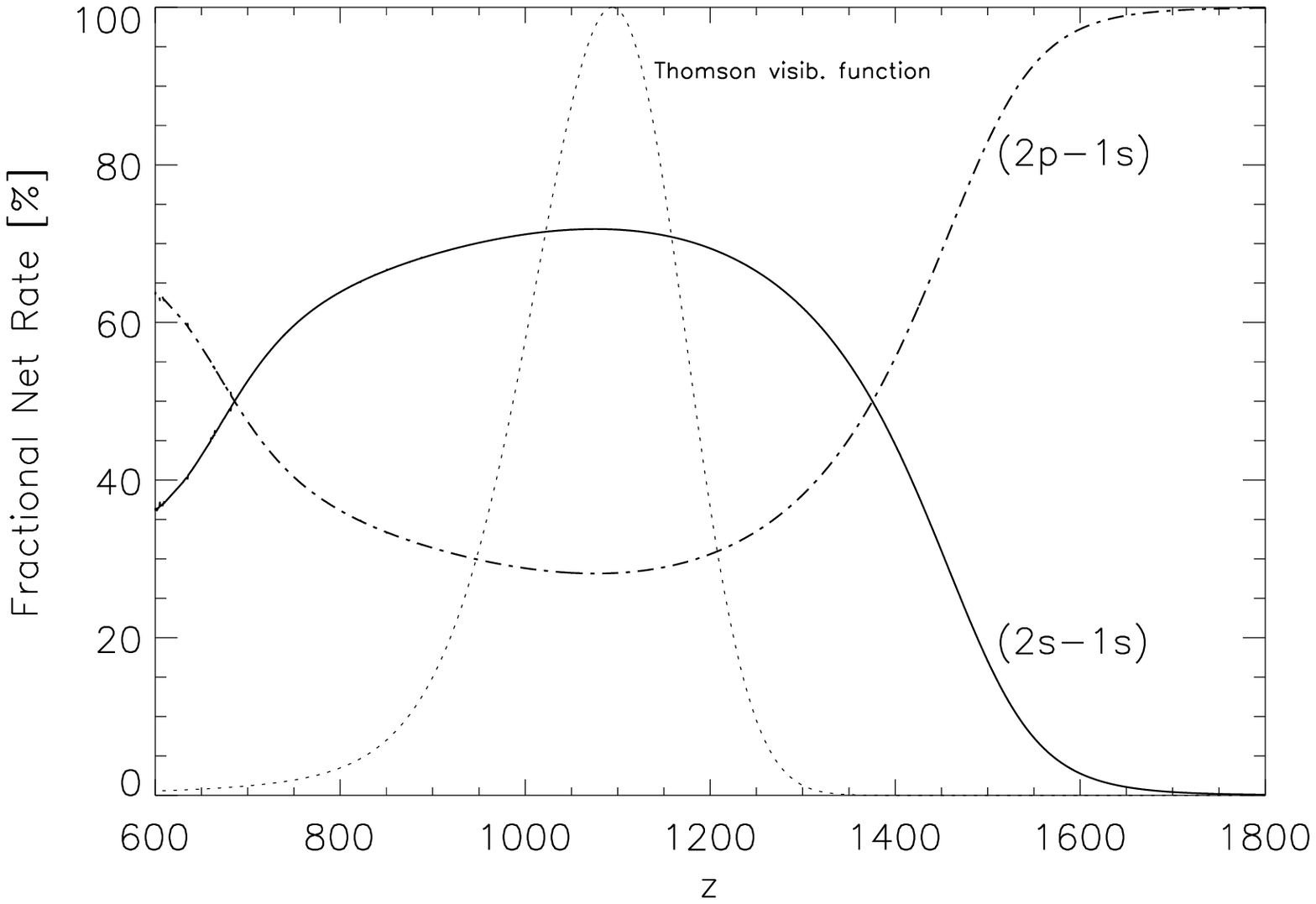}

\caption{\rm Fractional net rates for the two transitions from levels $n=2$ to
  the ground state, computed using eq.~\ref{eq:frate}. Around the peak
  of the Thomson visibility 
  function, the  $\Delta R(2s,1s)$ rate roughly contributes to $\sim 2/3$ of
  the total rate, while the Lyman-$\alpha$ route takes the other $\sim
  1/3$. At higher redshifts, the Lyman-$\alpha$ route is the dominant
  recombinational channel. At redshifts below $z\sim 700$, this
  Lyman-$\alpha$ route becomes dominant again due to the higher
  statistical weight of the 2p level as compared to the 2s (ratio of 3
  to 1).  }
\label{fig:frates}
\end{figure}

In order to quantify the relative weight of the two routes, we define
the fractional rates for both transitions as
\begin{equation}
\label{eq:frate}
F_{\rm 2p,1s} = \frac{\Delta R(2p,1s)}{\Delta R(2p,1s)+\Delta R(2s,1s)}, \qquad
F_{\rm 2s,1s} = \frac{\Delta R(2s,1s)}{\Delta R(2p,1s)+\Delta R(2s,1s)}.
\end{equation}
Figure~\ref{fig:frates} showed those fractional rates in the same redshift range as in
Figure~\ref{fig:rates}. In this way, the relative contribution of 2s
and 2p channels is more explicit. It is also interesting to see that
the redshift range where $\Delta R(2s,1s)$ approximately coincides with the peak of
the Thomson visibility function.  

The relevant quantity to understand both curves is the ratio of the
two effective radiative rates, $R \equiv \Delta R(2p,1s) / \Delta
R(2s,1s)$. This function is shown in Figure~\ref{fig:R}. As it is well
known, at redshifts higher than those where the bulk of recombination takes place, the
Lyman-$\alpha$ rate is dominant; the 2s channel starts to
dominate only at  $z \la 1400$ (e.g. ZKS68).
The interesting behaviour appears at low redshifts: once the
recombination is basically completed ($z \la 700 $), the Lyman-$\alpha$
escape route starts to dominate again. The asymptotic behaviour of this
$R$ function in this redshift regime can be easily estimated from
eq.~\ref{eq:DRij}. If we approximate the Sobolev escape probability
for the Lyman-$\alpha$ photons as $1/\tau_S$, being
$\tau_S$ the Sobolev optical depth (see e.g. Seager, Sasselov \& Scott
2000), we have
\begin{equation}
R \equiv \frac{\Delta R(2p,1s) }{\Delta
R(2s,1s)} \approx \frac{ \frac{8\pi \nu_\alpha^3}{3c^3} H(z) N(2p)/N(1s) }{
  A_{2s,1s} N(2s) }
\label{eq:Rapprox}
\end{equation}
The numerator in the last term of the previous equation corresponds to the rate of photons passing below the Lyman-$\alpha$
threshold, while the denominator corresponds to the rate of generation
of 2s two-photon pairs. Once the recombination is almost completed ($z\la
1000$), we have $N(1s) \approx N_{\rm H}(z)$, and inserting $N_{\rm
  H}(z)$ in equation~\ref{eq:Rapprox} we obtain the dashed line in
Figure~\ref{fig:R} labeled as low-$z$ asymptotic. This simple
equation can be used to estimate the redshift at which $\Delta
R(2p,1s)$ and $\Delta R(2s,1s)$ become comparable. As stated above,
this crossing occurs around $z \sim 700$. 

It is interesting to note that at those low redshifts, the relative ratio of the
population of the 2s and 2p levels is significantly out of Boltzmann
equilibrium, as shown in the numerical computations of RCS06 (see also
e.g. Chluba et al., 2007; and  Hirata \& Forbes 2009).  
For illustration, we also show in Figure~\ref{fig:R} the asymptotic behaviour
of equation~\ref{eq:Rapprox} if we ``incorrectly'' assume that the ratio
of the populations 2p:2s is given by the statistical weights, i.e., 3:1. In that
case, the equation~\ref{eq:Rapprox} reduces to 
\begin{equation}
R(SE) \approx \frac{ 8\pi \nu_\alpha^3 H(z) / c^3 }{  A_{2s,1s} N_{\rm
    H}(z) }.
\label{eq:Rapprox2}
\end{equation}
In the case of using this wrong assumption about the ratio of the 2p
and 2s populations, the estimate for the redshift at which the two rates $\Delta
R(2p,1s)$ and $\Delta R(2s,1s)$ become equal is artificially shifted towards lower values. 

For completeness, Figure~\ref{fig:n2p_n2s} explicitly shows the
computed ratio of the populations of the 2p and 2s levels, showing the
strong deviation from the equilibrium value of 3 at low redshifts.
These deviations originate as a consequence of the
recombination process, and in particular, due to the cascade of
electrons as recombination proceeds. In particular, we see that at low
redshifts, we have an overpopulation of the 2p level with respect to
the 2s. 
To understand this figure, it is also interesting to consider the
population of the levels in the $n=3$ shell. Figure~\ref{fig:n2n3} explicitly shows the
deviations from the statistical equilibrium for both the $n=2$ and
$n=3$ sub-levels, also using the same computations from RCS06. 
As the 2p level is directly connected via electric dipole
transitions with 3s and 3d, while the 2s level is only connected to
the 3p level, it is interesting to see how the non-equilibrium
conditions in the higher $n=3$ levels propagate down to the $n=2$
level. 
%


This work has been partially supported by the Russian Science
Foundation through grant 14-22-00271, and by the Spanish Ministry of Economy and
Competitiveness (MINECO) under the project AYA2014-60438-P. 
JAR-M acknowledges the hospitality of the MPA
during his visit. RS thanks V.~Mukhanov for his useful remark that
motivated this paper. We thank J. Chluba for useful comments. 
 
\pagebreak   

\clearpage

\centerline {\bf FIGURE CAPTIONS}
\vspace{1 cm} 

Fig.~\ref{fig:frates}.~Fractional net rates for the two transitions from levels $n=2$ to
  the ground state, computed using eq.~\ref{eq:frate}. Around the peak of the Thomson visibility
  function, the  $\Delta R(2s,1s)$ rate roughly contributes to $\sim 2/3$ of
  the total rate, while the Lyman-$\alpha$ route takes the other $\sim
  1/3$. At higher redshifts, the Lyman-$\alpha$ route is the dominant
  recombinational channel. At redshifts below $z\sim 700$, this
  Lyman-$\alpha$ route becomes dominant again due to the higher
  statistical weight of the 2p level as compared to the 2s (ratio of 3
  to 1).  

\begin{figure}[h]
\epsfxsize=19cm
\hspace{-2cm}\epsffile{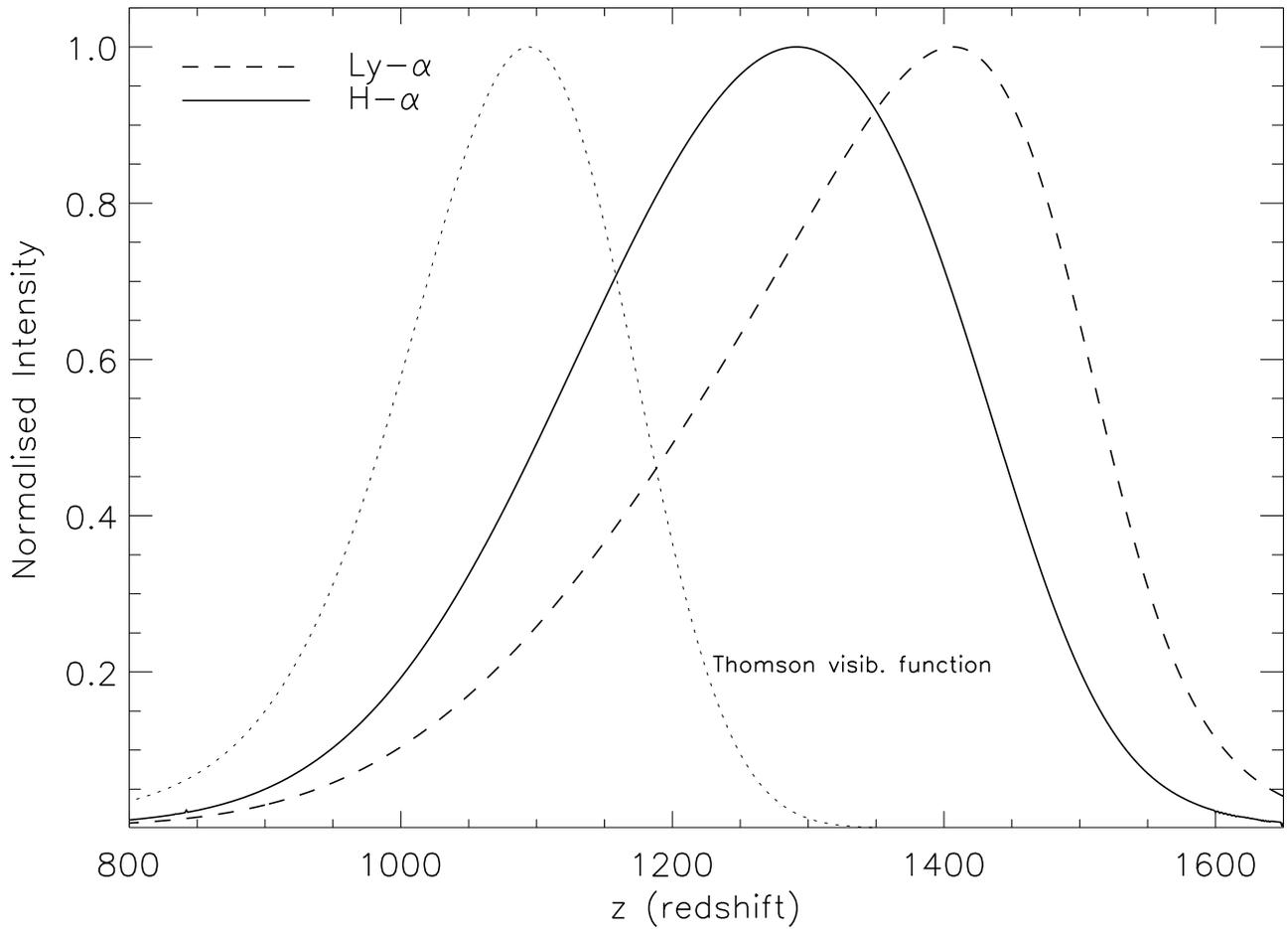}

\caption{\rm  Redshift of formation for the Lyman-$\alpha$ and H-$\alpha$ lines. We present the normalised
intensity for both transitions as a function of redshift z. For
comparison, we also show the Thomson visibility function normalised to unity at the peak, showing that the redshift of formation of the 
recombinational lines is higher than that of formation of the CMB
fluctuations. The low frequency wing of the Lyman-$\alpha$ line is
significantly suppressed due to increasing role of 2s two photon
decay. This is the main reason of the difference in the position of 
the maxima of the Lyman-$\alpha$ and H-$\alpha$ lines. Figure based from RCS06. }
\label{fig:zformation}
\end{figure}

\begin{figure}[h]
\epsfxsize=19cm
\hspace{-2cm}\epsffile{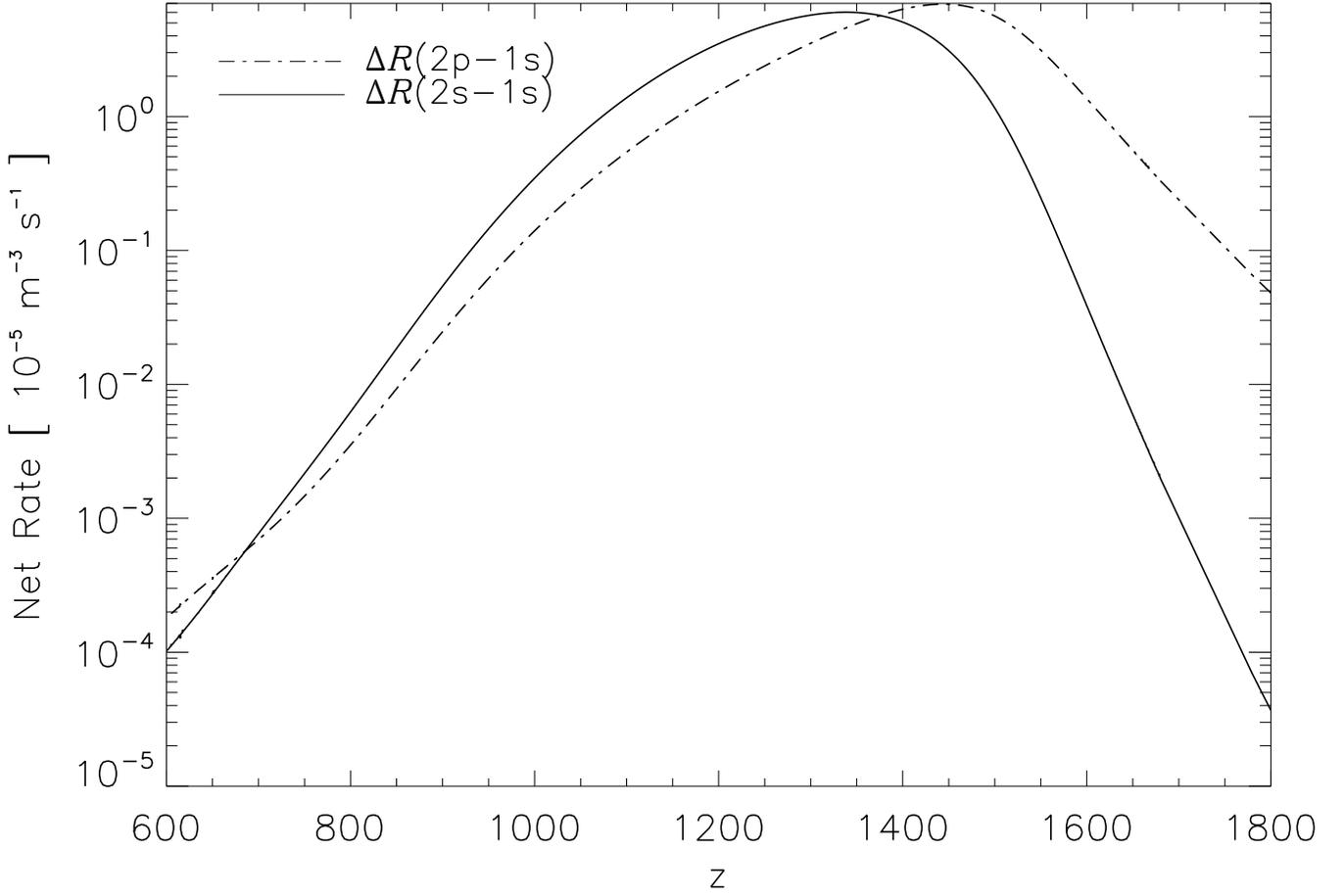}

\caption{\rm Net rates for hydrogenic transitions from levels $n=2$ to
  the ground state. The rate $\Delta R(2p,1s)$ mimics the shape of the
  Ly$\alpha$ line (to be compared with Fig.~9 in RCS06). Note that the
  2s two-photon decay rate dominates for redshifts between $z \sim
  700$ to $\sim 1400$, where the bulk of recombination is taking
  place.  }
\label{fig:rates}
\end{figure}

\begin{figure}[h]
\epsfxsize=19cm
\hspace{-2cm}\epsffile{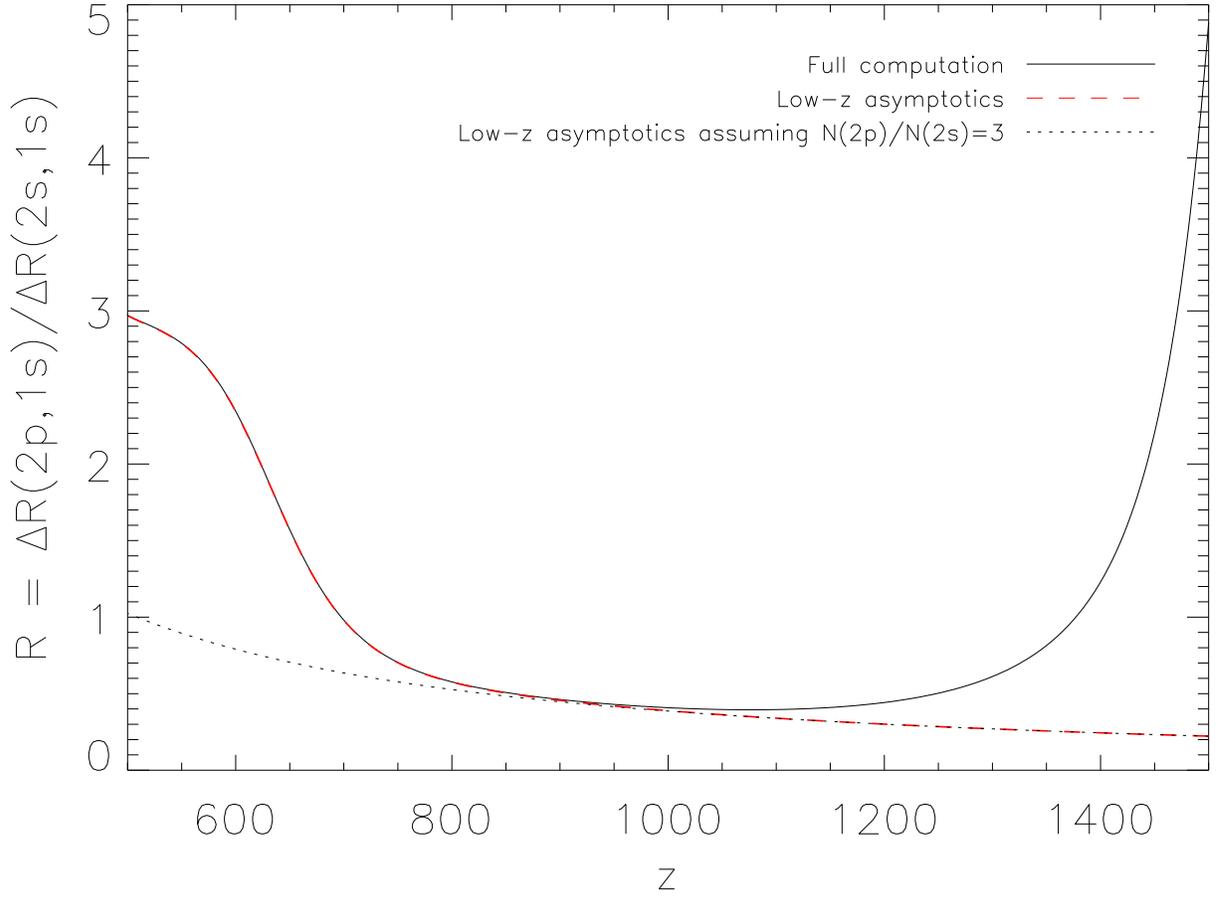}

\caption{\rm Ratio of the net rates $\Delta R(2p,1s)$ to $\Delta
  R(2s,1s)$ for hydrogenic transitions. For comparison, we also show
  as dashed-line the approximate asymptotic behaviour of this ratio as
  given by equation~\ref{eq:Rapprox}. The deviations from statistical
  equilibrium of the relative populations of the 2p and 2s levels are
  important to accurately compute the low redshift behaviour of this
  function (see text for details). }
\label{fig:R}
\end{figure}

\begin{figure}[h]
\epsfxsize=19cm
\hspace{-2cm}\epsffile{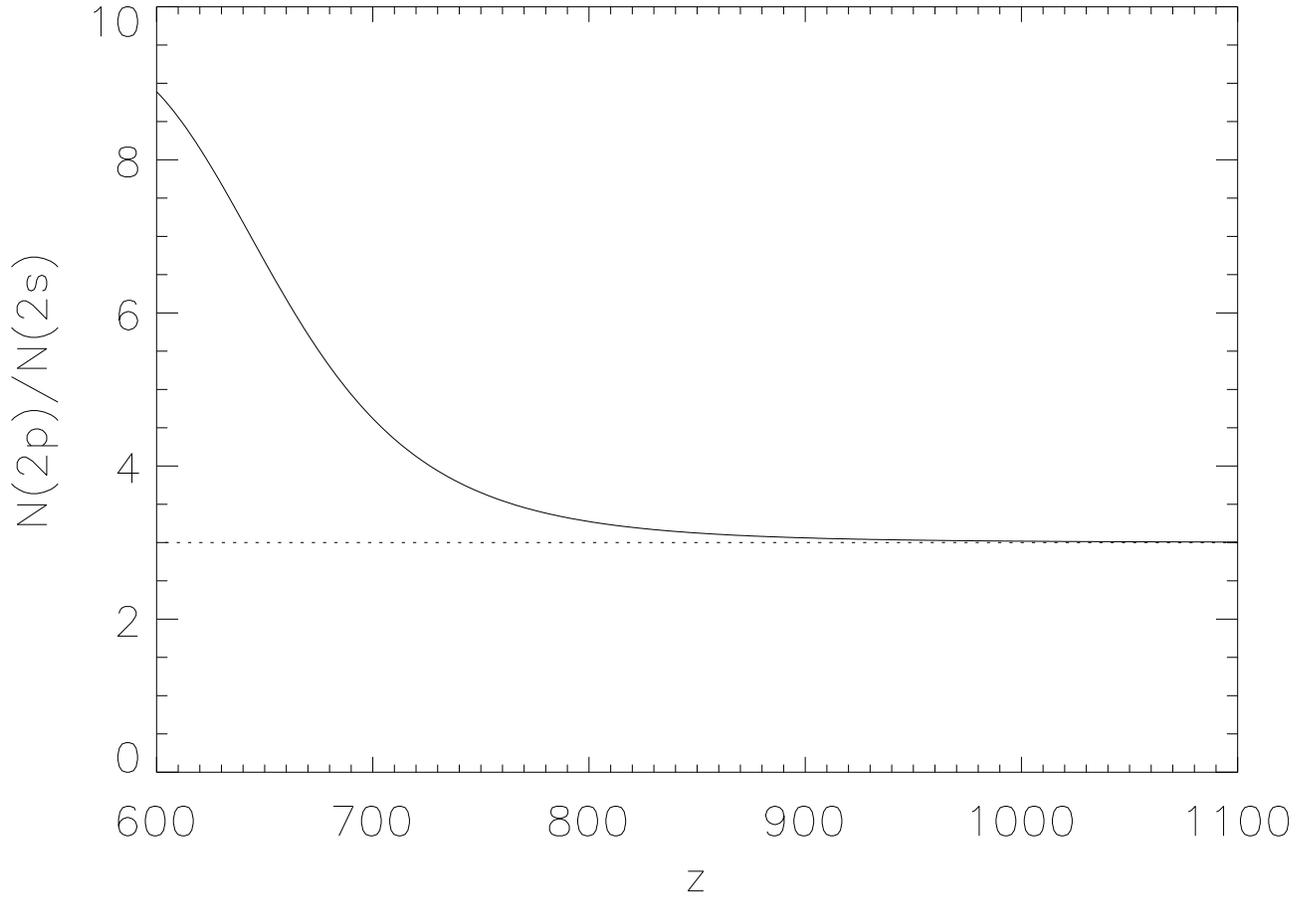}

\caption{\rm Ratio of the populations of levels $N(2p)$
  and $N(2s)$, as derived from the numerical code of RCS06. At low
  redshifts, this ratio strongly deviates from the statistical
  equilibrium ratio of 3:1 (horizontal dotted line). 
 }
\label{fig:n2p_n2s}
\end{figure}

\begin{figure}[h]
\epsfxsize=19cm
\hspace{-2cm}\epsffile{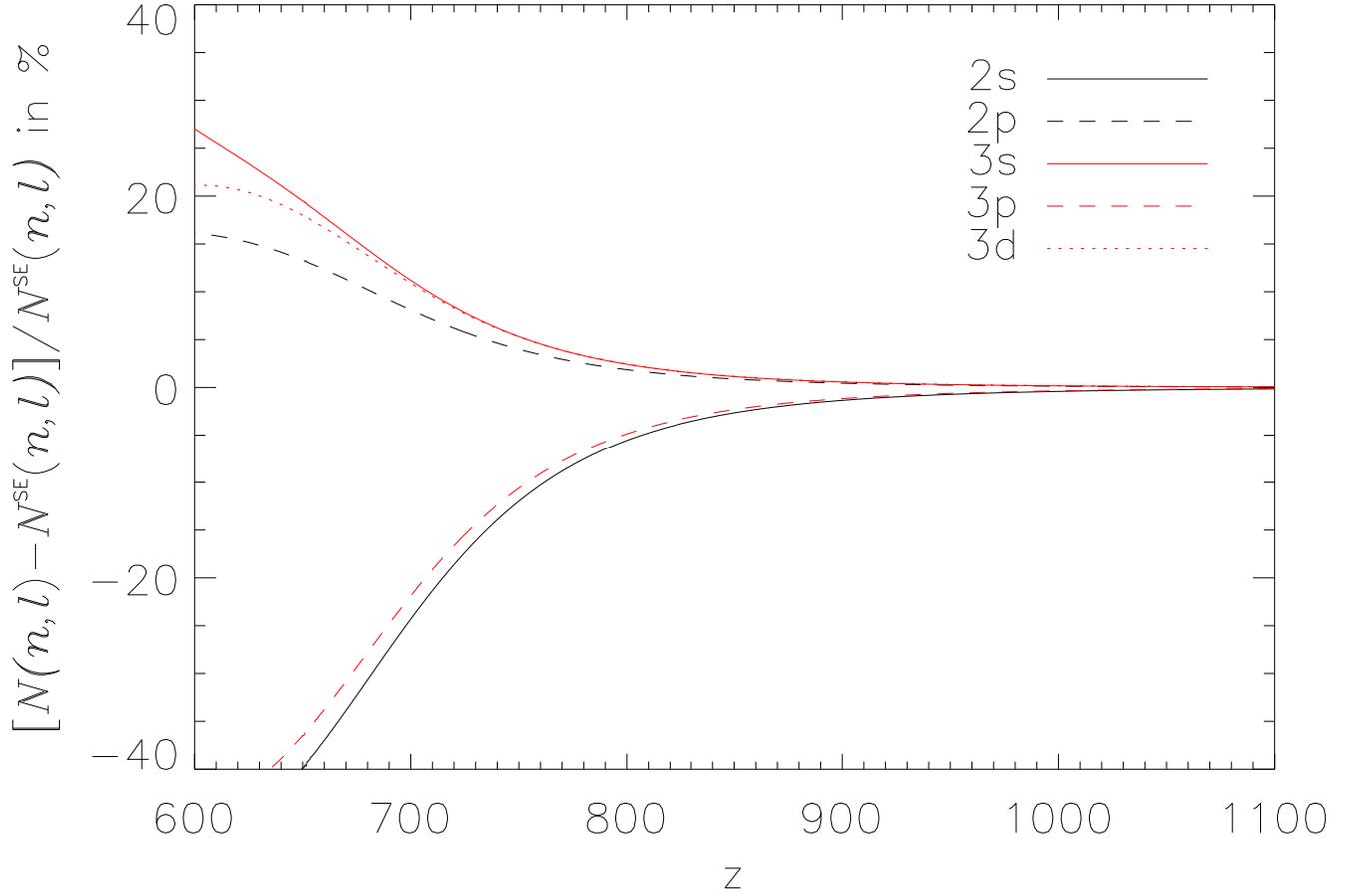}

\caption{\rm Non-equilibrium effects on the populations of the angular
  momentum substates for levels $n=2$ and $n=3$. We present the ratio
  $N(n,l)/N^{\rm SE}(n,l)$, where the SE index refers to the statistical
  equilibrium population, and is computed from the actual total
  population of the shell using $N^{\rm SE}(n,l) = [(2l + 1)/n^2] N_{tot}(n)$. 
 }
\label{fig:n2n3}
\end{figure}

Fig.~\ref{fig:zformation}.~Redshift of formation for the
Lyman-$\alpha$ and H-$\alpha$ lines. We present the normalised 
intensity for both transitions as a function of redshift z. For
comparison, we also show the Thomson visibility function normalised to
unity at the peak, showing that the redshift of formation of the  
recombinational lines is higher than that of formation of the CMB
fluctuations. The low frequency wing of the Lyman-$\alpha$ line is
significantly suppressed due to increasing role of 2s two photon
decay.  This is the main reason of the difference in the position of
the maxima of the Lyman-$\alpha$ and H-$\alpha$ lines. Figure based from RCS06. 

Fig.~\ref{fig:rates}.~Net rates for hydrogenic transitions from levels $n=2$ to
  the ground state. The rate $\Delta R(2p,1s)$ mimics the shape of the
  Ly$\alpha$ line (to be compared with Fig.~9 in RCS06). Note that the
  2s two-photon decay rate dominates for redshifts between $z \sim
  700$ to $\sim 1400$, where the bulk of recombination is taking place. 

Fig.~\ref{fig:R}.~Ratio of the net rates $\Delta R(2p,1s)$ to $\Delta
  R(2s,1s)$ for hydrogenic transitions. For comparison, we also show
  as dashed-line the approximate asymptotic behaviour of this ratio as
  given by equation~\ref{eq:Rapprox}. The deviations from statistical
  equilibrium of the relative populations of the 2p and 2s levels are
  important to accurately compute the low redshift behaviour of this
  function (see text for details).

Fig.~\ref{fig:n2p_n2s}.~Ratio of the populations of levels $N(2p)$
  and $N(2s)$, as derived from the numerical code of RCS06. At low
  redshifts, this ratio strongly deviates from the statistical
  equilibrium ratio of 3:1 (horizontal dotted line). 

Fig.~\ref{fig:n2n3}.~Non-equilibrium effects on the populations of the angular
  momentum substates for levels $n=2$ and $n=3$. We present the ratio
  $N(n,l)/N^{\rm SE}(n,l)$, where the SE index refers to the statistical
  equilibrium population, and is computed from the actual total
  population of the shell using $N^{\rm SE}(n,l) = [(2l + 1)/n^2] N_{tot}(n)$.

\clearpage


\clearpage

\clearpage


\clearpage

\clearpage

\clearpage


\end{document}